\documentclass[aps,superscriptaddress]{revtex4}
\usepackage{graphicx}

\begin{document}

%Title of paper
\title{Online games: a novel approach to explore how partial information influences human random searches}

% Alternative title: Human searching strategies as non-Markovian self-adaptive processes: an approach based on online games.

\author{Ricardo Mart\'inez-Garc\'ia}
\email{ricardom@princeton.edu}
\affiliation{Department of Ecology and Evolutionary Biology, Princeton University, Princeton, NJ 08544, USA}

\author{Justin M. Calabrese}
\affiliation{Smithsonian Conservation Biology Institute, National Zoological Park, Front Royal, VA 22630, USA}
\affiliation{Department of Biology, University of Maryland, College Park, MD 20742, USA}

\author{Crist\'obal L\'opez}
\affiliation{IFISC, Instituto de F\'isica Interdisciplinar y Sistemas Complejos (CSIC-UIB), E-07122 Palma de Mallorca, Spain.}

\date{\today}
% \pagewiselinenumbers

\begin{abstract}

Many natural processes rely on optimizing the success ratio of a search process. We use an experimental setup consisting of
a simple online game in which players have to find a target hidden on a board, to investigate the how the rounds are influenced by the detection
of cues. We focus on the search duration and the statistics of the trajectories
traced on the board. The experimental data are explained by a family of random-walk-based models and probabilistic analytical
approximations. If no initial information is given to the players, the search is optimized for cues that cover an
intermediate spatial scale. In addition, initial information about the extension of the cues results, in general, in faster searches.
Finally, strategies used by informed players turn into non-stationary processes in which the length of each displacement
evolves to show a well-defined characteristic scale that is not found in non-informed searches.

\end{abstract}

\maketitle

\section*{Introduction}

The problem of searching for targets whose location is unknown arises in many fields and at different scales \cite{MendezChap6,Benichou2011,kagan2015search}.
Numerous examples appear in the natural sciences including in ecology \cite{Viswanathan2011,Viswanathan1999,MendezChap9,bartumeus2005animal,edwards2007revisiting}, 
biochemistry  \cite{gorman2008, Kantslere02403,bonnet2008sliding} and chemistry \cite{Haangi1990}. In addition, many human activities 
involve situations where a target has to be found. Some instances are the location of a lost object, rescue operations, 
or fugitive prosecutions \cite{frost2001review}. More recently, the development of eye-tracking technology has allowed the 
study of visual searches on screens \cite{najemnik2005optimal,credidio2012statistical,amor2016persistence}.
 In order to understand the social, biological and physical mechanisms behind these processes, it is essential to have empirical evidence 
of the performance of different strategies and how they are affected by environmental cues, regardless of whether they are employed by humans,  animals or bacteria \cite{levin1992problem}. 
Such data are also required to verify the mathematical models that have been proposed 
\cite{bartumeus2002optimizing,hein,Benichou2005,Chupeau2015,Vergassola2007,campos2015optimal,abe2015levy,Robertse12572}, and to develop
improved protocols.

Situations in which a target has to be located appear in a large variety of scenarios, which allows the design of multiple strategies to find a successful solution.
Such strategies can be classified in many different ways, according to one or more of their properties \cite{MendezChap6}. For instance, stochastic or systematic processes are distinguished 
depending on the type of search rule \cite{Benichou2011} and the amount of directional information available determines the existence of bias towards preferred regions \cite{patlak1953random,codling2008random}.
Finally, differences
may also be attributable to the movement pattern, such as cruising versus ambush \cite{o1990search} and to the 
frequency of the reorientation events, such as intensive (frequent) versus extensive (infrequent)
\cite{JonsenEtAl2005.Robust.modeling,McClintockEtAl2012.Movement.framework}.
The effectiveness  of a particular choice within each category is determined by the properties and the state of the searcher,
the target, and the environment where the task has to be accomplished. 
For instance, searchers with memory that navigate relatively predictable environments do not employ purely random strategies but combine a stochastic component with 
knowledge acquired through previous experience. There is therefore a learning
process that plays an important role in the emergence of new rules \cite{merkle2014memory,fagan2013spatial}.
In other scenarios, individuals who live in groups may incorporate information gathered by
conspecifics with their own in order to improve foraging efficiency. It has been recently showed that intermediate combinations between both types of cues result in 
more efficient searches regardless of the nature of the mobility pattern \cite{Martinez-Garcia2014a} and the spatial distribution of
the targets \cite{bhattacharya2014collective,Martinez-Garcia2013b}. However, the precise optimal balance between social and individual information is determined by each 
specific setup.

In all of these scenarios, interactions with the environment provide the searcher with information that may alter the effectiveness of a given strategy 
over the course of the search. Therefore, in the most general case, search strategies must be interpreted as dynamical processes consisting of several components rather than fixed
procedures. For instance, many predators respond to the detection of cues indicating the proximity of prey by increasing their turning angles
and reducing their speed in order to scan the local environment more carefully \cite{hassell1978dynamics,curio2012ethology}, which leads to 
concentration of the search activity in areas of high prey density \cite{kareiva1987swarms}. This behavior has been 
reported in several species of insects \cite{kareiva1986patchiness}, seabirds \cite{weimerskirch2007does,fauchald2003using} and also in human searchers looking for hidden resources in
open environments \cite{hills2013adaptive}. Other phenomena that trigger sudden changes in individual movement behavior are changes between habitats \cite{Ovaskainen2004.Diffusion.model} and changes in the amount
and quality of information gathered by the searcher \cite{bartumeus2005animal}.

In this work we propose the use of computer games as a new experimental approach with which it is possible to address
these and related questions in humans. This is particularly intriguing since, due to their cognitive abilities,
individuals might show a large diversity of complex responses to the same stimulus.
Despite substantial efforts aimed at understanding the theoretical concepts behind many search processes, 
a reliable and unifying empirical framework in which these ideas may be tested is still lacking.
The family of games presented here is a good candidate to fill this gap, as they can be accessed
online by a large number of players. This results in the generation of large and clean datasets. 
In addition, the rules and setup of the game can be experimentally manipulated so that different
mechanisms or strategies can be rigorously tested.  
Firsy, we address several questions related to search efficiency and investigate how the strategies change due to 
the amount and the quality of information acquired by the player at different stages of the game.
In a second step, the main features of these patterns are extracted from the data and used to develop a family of random walk models that
can be applied to predict human search behavior in other configurations of the game. 
The variety of experiments shown in this work reinforces the flexibility of our approach and aims to open a new route for the study of search problems. 

In the following section, after presenting the characteristics of the game, we show the empirical results obtained from two different setups.
In the first case, players have no information about the configuration
of the board, whereas in the second study they are provided with partial information about it.
Then, we formulate a family of models that capture the main mechanisms behind the experimental results and derive analytical
approximations to show the robustness of the results. Finally, all the previous steps are combined to develop a comprehensive framework in which it is possible to predict the optimal configuration of the landscape that yields faster 
searches. The paper finishes with a discussion of the results 
and opportunities for new lines of research.

\section*{Results}

\subsection*{Experimental setup}
\begin{figure}[ht]
\centering
 \includegraphics[width=0.7\textwidth]{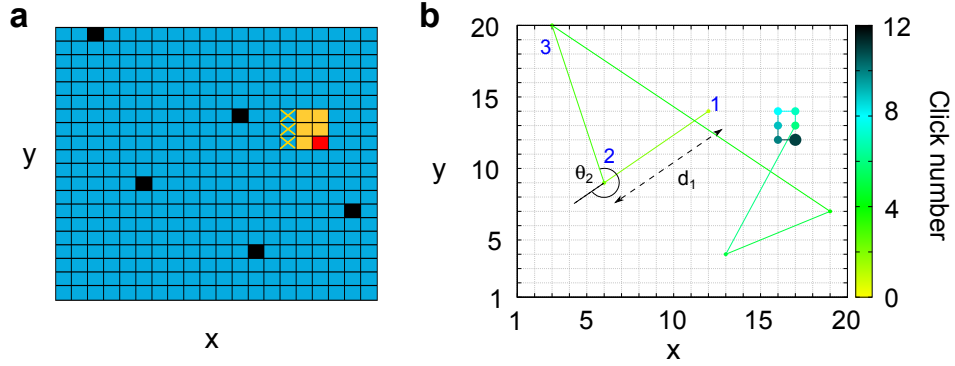}%
\caption{\label{fig:setup}  Experimental setup.  a) Single realization as shown in the game interface. Blue cells have not been visited,
black and yellow cells represent the two types of cues and the red square is the target. 
Yellow crosses mark those squares that belong to the neighborhood of the target and have not been visited yet. They are used here to indicate the layout of the board but they are not shown to the player.
b) Reconstruction of the round in A from the saved data. Small circles correspond to black 
cells, bigger circles to the yellow ones and the biggest circle is the target. Circles are labeled with blue numbers, $d_i$ is the distance
jumped starting from node $i$ and $\theta_i$ is the turn angle relative to the direction at node $i$.}
\end{figure}

We consider a simple game in which a single target has to be found.
It slightly resembles the classic {\it minesweeper}, although the objective is to find a unique target ({\it mine}) 
instead of avoiding a collection of them. The interface consists of $N\times N$ squares that can be explored 
by the player through successive clicks with the mouse. 
There are three classes of cells depending on their color after being clicked (unclicked cells are always blue): (i) black cells are typically far from the target,
(ii) yellow cells indicate that the target may be one of the neighboring cells and (iii) the single red cell is the target.
The target is randomly located within a patch of yellow cells. Therefore, it provides partial information about the configuration of the board.
Two different geometries for this set of yellow cells are explored here. First,  in the next two sections  they form
a $N_{\mbox{\tiny{y}}}\times N_{\mbox{\tiny{y}}}$ {\it neighborhood} square region (Fig.~\ref{fig:setup}a). Second,  in the last section of the Results  they will outline a random patch whose size will be measured in terms 
of the number of yellow cells.  Further details about the implementation of these random neighborhoods will be provided in that section. 
The discovery of a yellow cell indicates that the player is in the neighborhood of the target and thus reduces the area that needs to be scanned. 
For simplicity we fixed $N=20$ in all the experiments and then manipulated $N_{\mbox{\tiny{y}}}$.

To generate the dataset players access anonymously the game online and
are asked to find the target using as few clicks (jumps on the board) as possible.
Since players are not identified separately, we cannot identify the number of rounds played by each user.
The rounds are all independent (different configurations of the board) and each one 
is represented by the trajectory traced by the player on the board.
Finally, the experimental setup also includes a timer.  In order to study the limit in which searches are more stochastic, players are requested to find the target as quickly as possible.
This constraint also mimics many real situations both for humans and non-humans in which time is a limitation for the search.
Some instances are human rescue operations or animal foraging while avoiding predators.  

In the following sections we investigate i) how
the duration of the search, represented by the number of mouse clicks, changes with the size of the target's neighborhood (also called yellow region);
and ii) the statistical properties of the searching patterns as defined by the distance between clicks $d_i$ (jump length)
and the turn angles $\theta_i$. By definition, we consider turns to the left to be between $0^{\circ}$ and $180^{\circ}$ 
and turns to the right to be between $180^{\circ}$ and $360^{\circ}$ (see Fig.~\ref{fig:setup}b for a definition of both quantities).
We consider two classes of experiments: a) blind searches, where the player is given no \textit{a priori} knowledge of 
the size of the neighborhood, and b) searches with initial information, where the value of $N_{\mbox{\tiny{y}}}$ is given to the player at the 
beginning of the round. The objective of performing both classes of experiments is twofold: on the one hand to investigate whether players 
adapt their searching strategies when they have better information about the landscape and, on the other hand, to examine how search
efficiency changes when the reliability of the information provided by the yellow cells increases.

\subsection*{Experiments with blind searchers  and square neighborhoods }

\begin{figure}[ht]
\centering
 \includegraphics[width=0.78\textwidth]{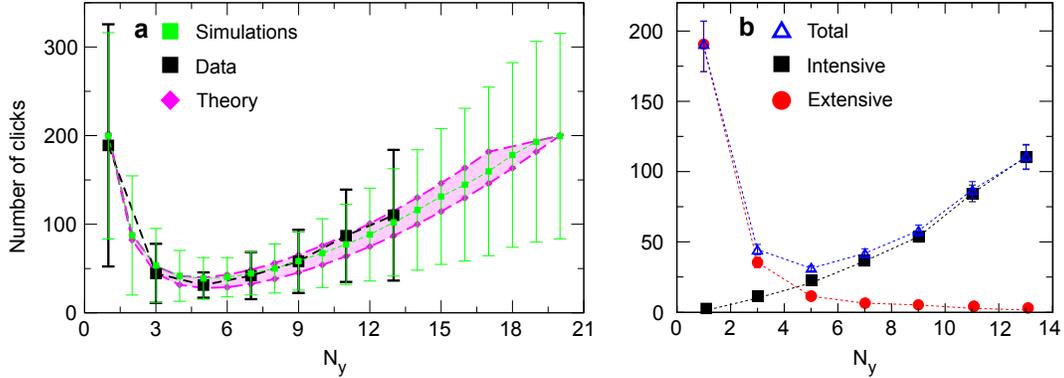}%
\caption{\label{fig:blindjumps} Number of movements for the blind searches as a function of the lateral length of the yellow neighborhood $N_{\mbox{\tiny{y}}}$.
a) Data-model-theory comparison of the total search length. $N_{\mbox{\tiny{y}}}=1$ means that there are no yellow cells around the target. 
Black squares are averages taken from experimental data, light green squares are obtained from numerical simulations (averages over
$10^4$ independent realizations) and the magenta region is the theoretical approximation. Dashed lines are interpolations and the error bars represent the standard deviation of the data. b) Decomposition of the 
total number of clicks between the intensive and the extensive phase. Dashed lines are interpolations and the error bars represent the standard error. When the bar is not shown the error is lower than the size of the point.}
\end{figure}

For this first series of experiments neither the exact
size, the position of the yellow region, nor a range of possible
dimensions was given to the searchers. Before starting the round, each player only knew 
that a target (red square) was hidden in the board and it might be randomly placed inside a square vicinity of
yellow cells of unknown size. The uncertainty in the size of the neighborhood reduces
the reliability of the information acquired by the player when a yellow cell is clicked and favors the efficiency of 
random strategies \cite{MendezChap6}. Our dataset consists of $500$ rounds with $N_{\mbox{\tiny{y}}}$ ranging from $N_{\mbox{\tiny{y}}}=1$,
which means that the target does not have a neighborhood, to $N_{\mbox{\tiny{y}}}=13$.  A distribution of the number 
of rounds for each value of $N_{\mbox{\tiny{y}}}$ is shown in the Supplementary Table I. 
We first measure the mean number of clicks needed to find the target as a function of the
lateral length of its yellow neighborhood (black squares in Fig.~\ref{fig:blindjumps}a).

Due to the design of the experiments, there is a tradeoff between 
finding the yellow region and finding the target inside it. 
Larger neighborhoods are easier to locate but make
 the final detection of the target inside them harder.
Smaller neighborhoods, however, need on average more steps to be found 
but make 
the target within them easier to locate (Fig.~\ref{fig:blindjumps}b). According to our results,
this tradeoff is balanced at intermediate sizes 
of the neighborhood, $N_{\mbox{\tiny{y}}}^{\mbox{\tiny{opt}}}=5$. This resembles the foraging dynamics of animals
that exchange information about food location with their conspecifics, so  
that both spreading information over distances that are either too large or too short may slow down the search \cite{Martinez-Garcia2013b}.
Following this analogy, we refer to the the size of the yellow area that minimizes the number of clicks needed to find the target as the {\it optimal interaction range}.
The standard deviation of the number of jumps is also
minimal at the optimal range, which means a narrowing in the distribution of clicks used to detect the target 
and therefore a reduction in the stochasticity of the search.
In the limit of zero information (i.e. no yellow cells or 
$Ny=1$,  or the whole board is yellow, $Ny=N$), the probability of finding the target on the first click is given by the inverse of the 
number of available cells, $1/N^{2}$. In any subsequent movement, $m$, this probability is given by

\begin{equation}\label{eq:limitnoinfo}
 P_m = \frac{N^{2}-(m-1)}{N^{2}}\times\frac{1}{N^{2}-(m-1)},
\end{equation}

where the first term yields the probability of not having found the target in the previous $m-1$ clicks and the second term yields the probability of hitting the target once
$m-1$ squares have been visited. Equation (\ref{eq:limitnoinfo}) reduces to $1/N^{2}$ regardless of the value of $m$.
Therefore, the probability of detection in the limit $N_{\mbox{\tiny{y}}}=1$
(and $N_{\mbox{\tiny{y}}}=N$) follows a uniform distribution of mean $N^{2}/2$ 
and standard deviation $N^{2}/\sqrt{12}$, which is
in good agreement with data (black squares in Fig.~\ref{fig:blindjumps}).

\begin{figure}[ht]
\centering
 \includegraphics[width=0.48\textwidth]{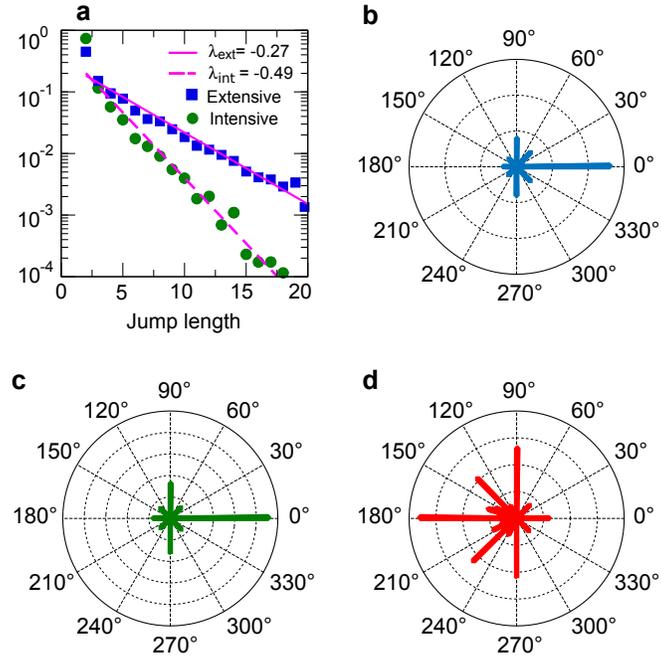}%
\caption{\label{fig:blindexp}  Statistical analysis of the trajectories on the board.  a) (Linear-log plot) Jump length distribution during the extensive (blue squares) 
and intensive (green circles) phase. Magenta lines are exponential
fits with mean value given by $1/\lambda$. b, c) Turn angle 
distributions during the extensive and the intensive phase respectively.
d)  Turn angle distribution for movements performed immediately after a yellow-to-black transition. }
\end{figure}

Next, we analyze all the trajectories traced by
the players in every round. To facilitate this, the
experimental setup saves the sequence of clicks in each round, from which we calculate the length of each displacement and the angle of each turn.
We identify extensive and intensive searching modes that depend on 
whether the player has detected a yellow cell or not respectively (Fig.~\ref{fig:blindexp}a).
In both situations the jump lengths can be fitted using exponential distributions,  with the intensive phase showing a lower mean value 
$1/\lambda_{\mbox{\tiny{int}}}=2.04$ and $1/\lambda_{\mbox{\tiny{ext}}}=3.70$ 
($1/\lambda_{\mbox{\tiny{int}}}$, $1/\lambda_{\mbox{\tiny{ext}}}$ are the mean length of the displacements in the intensive and the extensive phase respectively).
Therefore, the typical size of the jumps is reduced once the player finds the yellow area
as the detection of the cue (represented by a yellow cell)  triggers an area-restricted search \cite{hassell1978dynamics,curio2012ethology}.  
Although the player does not know how big the neighborhood is and therefore how reliable the information is, 
the trajectories recorded after the discovery of the yellow region still show shorter distances between turns, suggesting that players switch to an intensive search mode once they find the yellow region
\cite{MendezChap6}.  It is important to remark that, although alternation between extensive (motion phase) an intensive modes (scanning phase) is also characteristic
of intermittent searches, the player is not performing an intermittent search as it has been defined in the literature before \cite{Benichou2011}. The differences lie in two points. First, in our study
the switch between reorientation modes is triggered by the external cue instead of taking place at random. Second, the detection of the target may take place in both phases instead of being limited to 
the extensive one. Regarding to the type of motion, we study, however, a spatially intermittent search since the player performs a saltatory trajectory in which the target can be found only if 
the searcher lands on it. This differs from the case of a cruise forager who looks for targets while moving and that would constitute a completely different study. 

Regarding the turn angles, both the extensive (before the first encounter with a yellow cell) 
and the intensive phases (after detecting the first yellow cell) show correlations between subsequent turn angles (Fig.~\ref{fig:blindexp}b,c, respectively).
This could indicate that the strategies 
are not completely random but contain some systematic features.
In fact, a frequent strategy consists of tracing a series of short jumps
in the same direction. To reduce searching times players show a 
tendency to scan a direction doing several consecutive clicks. This behavior is also seen in the distributions of jump lengths,
since they show a large deviation from the exponential for one-cell 
length jumps, which are overrepresented in the dataset (Figure~\ref{fig:blindexp}a). 
The higher frequency of turning angles closer to zero is  linked
to the higher presence of jumps of length one. The explanation for this persistence in the direction of movement shown in Fig.~\ref{fig:blindexp}b,c is probably
a combination between the attempt of some players to design purely systematic strategies and 
the intrinsic tendency of humans to keep visually scanning in the same direction \cite{amor2016persistence}. 

As an exception, movements done immediately after a yellow-to-black transition show a 
strong tendency to reverse the direction, as this sequence in the colors of the cells indicates 
that the player is moving away from the target (Fig.~\ref{fig:blindexp}d).

\subsection*{Experiments with initial information  and square neighborhoods. The case of $N_{\mbox{\tiny{y}}}=5$ as compared to the blind case}\label{sub:know}

In this second series of experiments the players know the size of the yellow region, 
which is fixed at the optimal interaction range $N_{\mbox{\tiny{y}}}=5$. This increases the quality of the information obtained when one of the yellow
cells is found as the player can limit the search area. The position of this area, as well as the location of the target inside it, is random, changes from
round to round and is unknown to the player.

\begin{figure}[ht]
\centering
 \includegraphics[width=0.48\textwidth]{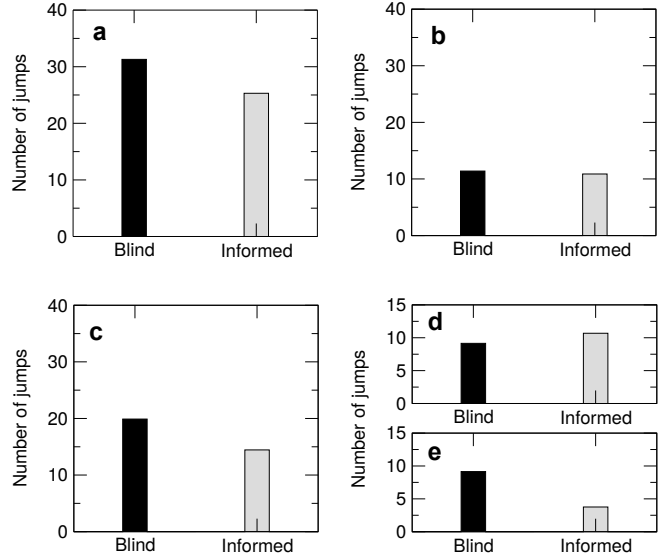}%
\caption{\label{split} Factorization of the number of clicks comparing blind and informed searches with $N_{\mbox{\tiny{y}}}=5$. a) Total number of clicks, b) extensive phase, c) intensive phase,
d) yellow-to-yellow jumps of the intensive phase, e) black-to-yellow and yellow-to-black transitions and black-to-black movements during the intensive phase.}
\end{figure}

Data from $230$ rounds were collected. As a general result, \textit{a priori} information accelerates 
the search and reduces its stochasticity. Blind searchers need on average $31.30$ 
clicks to find the target when $N_{\mbox{\tiny{y}}}=5$ (subset of $65$ rounds from the $500$ trajectories 
analyzed for blind experiments, see  Supplementary Table I ), while  informed  players use
$25.5$ clicks. The two-tailed P value on the difference of these mean values obtained using an unpaired t-test, $3\times10^{-4}$,
is highly statistically significant. The standard deviation also decreases, indicating a narrowing
in the distribution of the number of displacements and therefore in the randomness 
of the process: $\sigma_{\mbox{\tiny{b}}}=14.10$ for blind searchers and $\sigma_{\mbox{\tiny{i}}}=10.50$ for the informed ones.
 To find out what stage of the search is more strongly affected by the initial information,  we analyze 
the number of clicks done in each phase of the search. We repeat this factorization for the blind and the informed cases and 
compare both of them (Fig.~\ref{split}). From Fig.~\ref{split}b and \ref{split}c,
we observe that all of the reduction in the number of jumps accumulates in the intensive phase, while the extensive stage remains unaltered by the initial information.
More interestingly, if the number of displacements that take place between two yellow cells is subtracted from the total number of jumps of the intensive phase (Fig.~\ref{split}d),
we observe that this quantity remains almost unchanged. There is, however, an important reduction in the number of displacements that correspond to the rest of the combinations of cells
(black to yellow, yellow to black and black to black jumps; Fig.~\ref{split}e). In fact, the percentage of yellow-to-yellow movements that take place during the intensive phase increases from $54\%$ to
a $75\%$ in the informed searches. This result indicates that having information about the size of the yellow zone allows
a faster detection of its limits and therefore reduces the number of movements spent to find the target.  

\begin{figure}[ht]
\centering
 \includegraphics[width=0.48\textwidth]{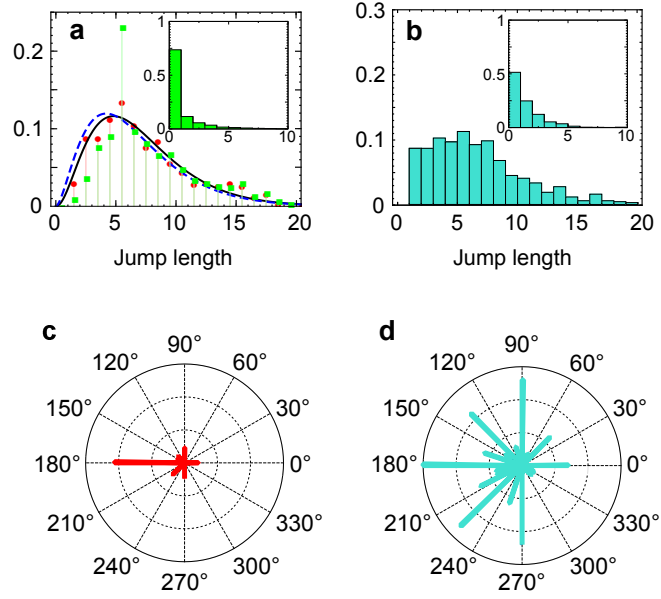}%
\caption{\label{fig3}  Comparison of the jump length and turning angle distributions for informed and blind searchers with $N_{\mbox{\tiny{y}}}=5$ . a) Jump length distribution 
for the extensive phase.  Green squares
correspond to the whole set of rounds and red circles to the subset of random strategies.
The dashed and full lines show two analytical approximations. Inset: distribution for the intensive phase. b) Equivalent to a) but for the subset of $65$ blind
rounds with $N_{\mbox{\tiny{y}}}=5$. c) Turning angle distribution for movements in the intensive phase made immediately after a yellow-to-black jump.
d) Same as c) but for the subset of blind searches with $N_{\mbox{\tiny{y}}}=5$.}
\end{figure}

 Regarding to the statistical analysis of the trajectories, initial information about $N_{\mbox{\tiny{y}}}$ also yields some differences in the distributions
of the lengths of the jumps and the turning angles. Informed players adapt their displacements during the extensive phase, concentrating 
the length of their movements around the size of the yellow neighborhood, $N_{\mbox{\tiny{y}}}=5$ (Fig.~\ref{fig3}a). 
If we analyze the whole set of  informed  rounds, we observe a strong dominance of movements of length $N_{\mbox{\tiny{y}}}=5$ (green squares in Figure~\ref{fig3}a).
This is due to the presence of approximately $50$ rounds in the dataset
where players performed optimally designed systematic strategies that consist of moving in jumps of fixed length $N_{\mbox{\tiny{y}}}$ during the extensive phase.
We will come back to this in the description of a random walk based model for this process. For the purposes of this section we will remove these systematic rounds and focus on the
subset of stochastic strategies formed by the other $180$ rounds. The distribution of the length of the displacements is still dominated by jumps
that cover a distance of the order of $N_{\mbox{\tiny{y}}}$ (red circles in Fig.~\ref{fig3}a). For the subset of blind searchers with $N_{\mbox{\tiny{y}}} = 5$, however, this distribution does not show a well defined typical
scale and instead, players explore several scales as they look for the yellow region (Fig.~\ref{fig3}b).
For the intensive phase,  informed  searches also show a higher abundance of one-cell displacements than the distribution of the blind 
searches (inset of Fig.~\ref{fig3}a and \ref{fig3}b respectively). This result is independent of whether or not the systematic deterministic strategies are included within the analyzed dataset and 
is due to the fact that knowing the neighborhood size reduces exploration during this phase. Finally, giving the size of the yellow neighborhood to the players in advance also has an effect on the distribution 
of turns made by the searcher immediately after a yellow-to-black displacement.
This distribution is shown in Fig.~\ref{fig3}c for informed strategies and in Fig.~\ref{fig3}d for blind searches. 
Although in both cases the movement shows a strong bias backwards, informed searches result in distributions with a stronger peak around $\theta = 180^{\circ}$.  This is due to the fact that
players do not have to find out the size of the neighborhood of the target and consistent with the factorization of the number of clicks shown in Fig.~\ref{split}

\begin{figure}[ht]
\centering
 \includegraphics[width=0.48\textwidth]{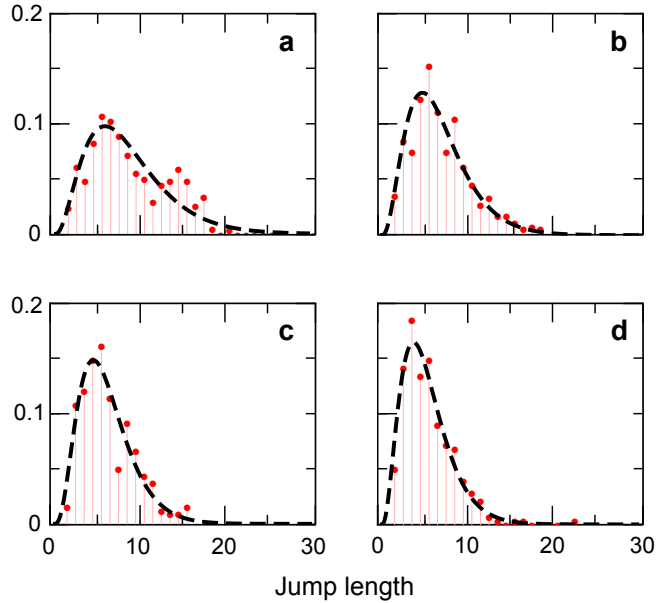}%
\caption{\label{nonst}  Non-stationary jump length distributions for the extensive phase of the informed searches .
The extensive phase is divided in four parts and the distribution of each subset is shown: steps 1-5 (a), 6-10 (b), 11-15 (c) and 16-end (d). Red circles show experimental data and black dashed lines the theoretical fitting. 
Parameter estimates for each fit are shown in Table \ref{tabla-partidas}.}
\end{figure}

 We conclude this section with an analysis of the trajectories during the extensive phase, in order to 
find the mechanism by which a characteristic length scale appears in the jump length distribution. We find the existence
a feedback between the searcher and the environment that makes the extensive phase non-stationary (the mean value of the distribution changes with time). This feedback allows 
a progressive narrowing of the jump length distribution around $N_{\mbox{\tiny{y}}}$ as the extensive phase evolves and the searcher gathers and accumulates information from the landscape.
Since the player has perfect memory about his trajectory (visited cells remain
open), trajectories that start with large displacements tend to create landscapes that are fragmented in patches of length $N_{\mbox{\tiny{y}}}$ in which long movements are inefficient.
To show the existence of this feedback we split the data of the extensive
phase in four subdivisions: (i) from jump 1 to 5, (ii) 6 to 10, (iii) 11 to 15, and (iv) 16 to the end. The distributions for each of these pieces are shown in Fig.~\ref{nonst}a, \ref{nonst}b, \ref{nonst}c 
and \ref{nonst}d respectively, and they can be fitted by a family of gamma distributions (dashed lines in each panel) of decreasing mean, mode and variance
(See Table \ref{tabla-partidas} for numerical values of these parameters and details of the distributions). 
Then the total distribution of Fig.~\ref{fig3}a can be approximated by a gamma function defined in terms of the parameters of the 
distributions of the pieces (dashed line in Fig. \ref{fig3}a). This approach shows an excellent agreement with a direct fitting of the whole extensive phase (full line in Fig. \ref{fig3}a). 

At this point, we have shown the existence of an optimal size for the neighborhood of the target, as well as an improvement in the search efficiency when
the value of $N_{\mbox{\tiny{y}}}$ is revealed at the beginning of the round. In addition, these informed strategies evolve through information gathering during the extensive phase 
towards a dominant jumping distance equal to the lateral length of the neighborhood of the target, $N_{\mbox{\tiny{y}}}$. In the following sections we develop a theoretical framework and a family of
models based on random walks to study the basic principles behind these results and how they can be transfered to more general scenarios, with irregular shapes for the information region.

\subsection*{Model for blind searchers: numerical simulations and analytical approximation}

We develop a minimalistic searching model 
based on random walks to explain previous experimental results on the basis of simple dynamical rules.
The model has the three main ingredients obtained from the data
analysis: (i) two modes of movement defined
by the mean length of the displacements: $1/\lambda_{\mbox{\tiny{int}}}$ and $1/\lambda_{\mbox{\tiny{ext}}}$; 
(ii) in the absence of any information (no yellow cell clicked) 
the direction is completely random (uniform distribution
in the turning angles); 
and (iii) when cues are obtained (a yellow cell has been detected), 
the searcher has a bias towards unvisited cells surrounding a yellow one. 
The choice of a uniform distribution for the turning angles is a consequence of using a purely exponential distribution for the 
length of the displacements (see Fig.~\ref{fig:blindexp}a). The high persistence shown by the experimental turning angle distribution, which can be approximated by a uniform distribution
except for that peak at $\theta = 0$ (see Fig.~\ref{fig:blindexp}b,c), comes from the high presence of jumps 
of length one. Disregarding the high frequency of unity-length movements also implies disregarding the higher abundance of turning angles close to zero and therefore using a uniform distribution for $\theta$.
 The third assumption aims to capture the influence of the information 
provided to the searcher when a yellow cell is found, as well
as the strong tendency to go back to yellow cells exhibited by the
distribution of turning angles in Fig.~\ref{fig:blindexp}d. 
% More details of the model as well as the simulation setup are provided in the Methods section.

The results of the simulations (green curve in Fig.~\ref{fig:blindjumps}) show an excellent agreement 
with the experimental data (black curve) both in the mean average number of jumps and in its standard deviation.
Simulations reproduce at least the two first moments of the number of clicks distribution.

Except in the limits $N_{\mbox{\tiny{y}}}=1$ (no yellow cells) and $N_{\mbox{\tiny{y}}}=N$ (yellow cells occupy the whole board),
it is hard to obtain exact analytical expressions for the average total number of jumps needed to find the target. 
However, it is possible to obtain the distribution for the length of the extensive phase

\begin{equation} \label{eq:extensive}
 P_i(N_{\mbox{\tiny{y}}}) = p_i(N_{\mbox{\tiny{y}}})\prod_{j=1}^{i-1}\left(1-p_j(N_{\mbox{\tiny{y}}})\right),
\end{equation}

where $P_i(N_{\mbox{\tiny{y}}})$ is the probability of having an extensive phase of $i$ jumps when the neighborhood of the target has a lateral length $N_{\mbox{\tiny{y}}}$
and $p_i(N_{\mbox{\tiny{y}}})=\frac{N_{\mbox{\tiny{y}}}}{N-i+1}$ is the probability of finding a yellow cell in the
$i-th$ mouse click. In words, the probability of having an extensive phase with $i$ jumps is given by the probability of not finding a yellow cell
in all the previous movements multiplied by the probability of finding one in the $i-th$ movement. Given Eq.~(\ref{eq:extensive}), the mean length 
of the extensive phase is 

\begin{equation} \label{eq:limits-anal}
M_{\mbox{\tiny{ext}}}=\sum\limits_{i=1}^{N^{2}-N_{\mbox{\tiny{y}}}^{2}+1}iP_i(N_{\mbox{\tiny{y}}}).
\end{equation}

For the length of the intensive phase however we can only give and
upper and a lower limit, assuming that after the detection of the first yellow cell all the movements are to neighboring cells. Therefore, the target is found on average after $N_{\mbox{\tiny{y}}}^{2}/2$
jumps in the intensive phase when the neighborhood of the target is large and after $(N_{\mbox{\tiny{y}}}+2)^{2}/2$ movements when the neighborhood is small. These two limits account for
the decreasing probability of visiting cells outside the neighborhood when increasing its size. For small values of $N_{\mbox{\tiny{y}}}$ it is very likely to reach the border of the 
neighborhood before detecting the target and thus to return to the black region. Combining these two results for the intensive phase with the length of the extensive phase
obtained in Eq.~\ref{eq:limits-anal}, we obtain two theoretical approximations to the total number of clicks

\begin{eqnarray}
 M^{\mbox{\tiny{up}}} &=& \sum\limits_{i=1}^{N^{2}-N_{\mbox{\tiny{y}}}^{2}+1}iP_i + \frac{(N_{\mbox{\tiny{y}}}+2)^{2}}{2}, \\
 M^{\mbox{\tiny{low}}} &=& \sum\limits_{i=1}^{N^{2}-N_{\mbox{\tiny{y}}}^{2}+1}iP_i + \frac{N_{\mbox{\tiny{y}}}^{2}}{2}.
\end{eqnarray}

The combination of these two expressions gives an approximated range for the length of the search (magenta region in Fig.~\ref{fig:blindjumps})
that shows an excellent agreement with empirical data and numerical simulations of the model. 

\subsection*{Model for searches with initial information. The design of optimal strategies.}

\begin{figure}[ht]
\centering
 \includegraphics[width=0.48\textwidth]{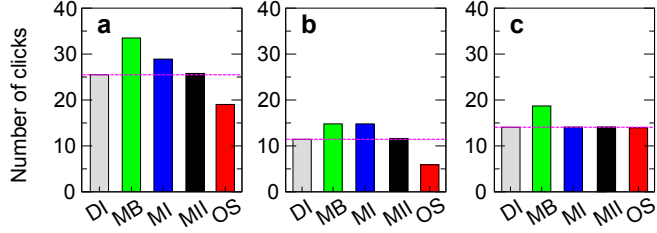}
\caption{\label{fig-comp} Comparison between informed searches experimental data and the models. a) Number of clicks before target detection (extensive + intensive).
b)  Number of clicks during the extensive phase. c) Number of clicks during the intensive phase. The magenta dashed line shows the value obtained from the data for informed searchers.
Labels of the x-axis: DI data informed, MB model blind, MI model informed, MII model informed 2 and 
OS optimal strategy.}
\end{figure}

Knowing the size of the yellow region at the beginning of the game changes the nature of the search as the information gathered by the player with each
movement may be used to design the next displacement. This reinforces the non-Markovian nature of the informed search process as the player uses all the previous steps to discard
cells that have not been visited yet and results in self-adaptive strategies that evolve towards displacements of length $N_{\mbox{\tiny{y}}}$. Also, as the value of $N_{\mbox{\tiny{y}}}$ is known,
the number of exits from the neighborhood of the target diminishes (Fig.~\ref{split}).
In a first approach to model this effect, we modify the model used 
for blind searches using the new experimental distribution of the length of the displacements in both the extensive and the intensive modes (Fig.~\ref{fig3}a).
Therefore, instead of using the exponential distributions of Fig.~\ref{fig:blindexp} we sample the histograms of Fig.~\ref{fig3}a (red circles) and its inset, 
that are obtained from experimental searches with initial information. 
This approach overestimates both the length of the extensive and the intensive phases, which results in a clearly higher average number of movements; $33.50$ jumps, $\sigma=19.00$ for the model
and $25.50$ jumps,  $\sigma=10.50$ in the data (MB green bars and DI gray bars in Fig.~\ref{fig-comp} respectively). This is due to the fact that the model does not integrate
the information about the size of the target to a priori discard some of the cells during the intensive and the extensive phase. 
% To unveil the role of each of this mechanisms we improve the modeling of the intensive phase first and then the description of the extensive stage.

In a first approach to remove this discrepancy, we hypothesize that the most important differences arise in the modeling of the intensive phase.
During this stage, given a certain number of yellow cells and some
of their neighboring black squares, our experimental results suggest that human players are able to discriminate the real border of the neighborhood of the target and thus
reduce the number of erroneous displacements. The model that we developed for blind searches lacks this ingredient, which increases the duration of the intensive phase since more black cells are open.
To correct this, we modify the model and include the effect that previous movements, together with knowing the size of the neighborhood of the target, have on the intensive
phase (See Methods for a detailed description). In this new approach, once the first yellow cell has been detected and based on all
the previous movements, only those cells that can possibly be part of a $5\times 5$ yellow square have
a non-zero probability of being visited by the searcher. 
This mechanism reduces the number of times that black cells are visited once a yellow cell has been found as the model is able to discriminate all the possible borders of
the neighborhood of the target. With this new ingredient
the efficiency of the model increases (MI blue bar in Fig.~\ref{fig-comp}a) and the number of jumps in the intensive phase shows excellent 
agreement with the experimental data (DI gray and MI blue bars in Fig.~\ref{fig-comp}c).  However, despite this substantial improvement 
as compared to the blind model, significant differences still remain between empirical data and numerical results.
The source of this disagreement arises from the extensive phase (DI gray and MI blue bars in Fig.~\ref{fig-comp}b). To correct this, we next modify the extensive phase of the model.

During the extensive phase, players are able to discriminate regions where the target cannot be placed as a $5\times5$ square would not fit.
To incorporate this in the model, we first compute the probability of jumping to each of the non-visited cells of the board
according to the histogram in Fig.~\ref{fig3}a. Then, for each cell we obtain all the possible squares of lateral length $5$ to which it could belong
and set the probability of jumping to that cell to zero if all these squares contain at least one open black cell (See Methods for more details).
With this mechanism the extensive phase becomes more efficient and the agreement of the model with the experimental data is excellent.
More importantly, this comes from a precise fitting of both the intensive and the extensive phase individually (DI gray and MII black
bars in Fig.~\ref{fig-comp}a,b,c).

{\it Optimal strategy.-} However, both actual player strategies and random walk models are much less efficient than entirely systematic protocols.
Knowing a typical size of the target in advance allows the design of optimized strategies that minimize the number of incorrect steps.
Particularly important is to shorten the extensive phase, as within the neighborhood of the target all the cells are equivalent
and it is equally likely to find the target in any position. In fact, during the experimental rounds with initial information, one of the players 
developed one of these searching methods by repeatedly playing with the same size of $N_{\mbox{\tiny{y}}}=5$.

\begin{figure}[ht]
\centering
 \includegraphics[width=0.48\textwidth]{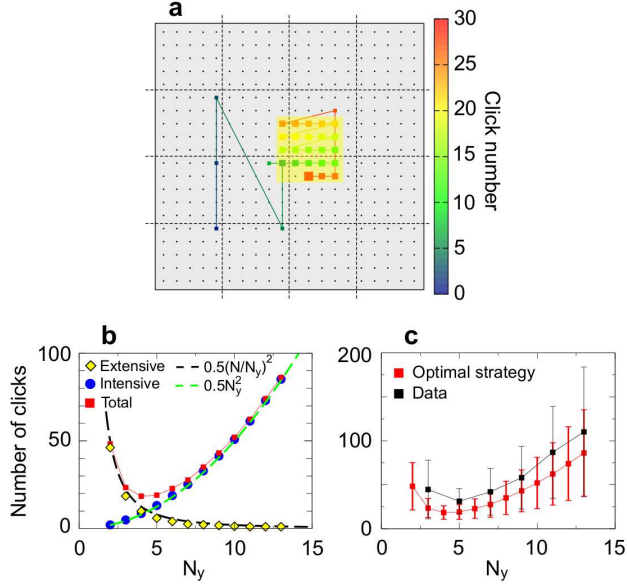}
\caption{\label{fig4} Analysis of the optimal systematic strategy. a) Typical realization. The color of the squares indicates the temporal sequence
of the jumps and its size the location outside (smaller squares) or inside the neighborhood (intermediate squares). The biggest square represents 
the target. b) Typical length of search as a function of the size of the neighborhood (red squares). This quantity is divided between the extensive (yellow diamonds) and
the intensive (blue circles) phases. Analytical approximations are shown by dashed lines.
c) Comparison between the mean number of jumps needed using an optimized systematic search rule (red squares) and the blind experimental data (black squares). Error bars represent the standard deviation, lines are interpolations.}
\end{figure}

This strategy optimizes the extensive phase and only allows for two yellow-to-black transitions during the intensive phase (Fig.~\ref{fig4}a).
Given a value for $N_{\mbox{\tiny{y}}}$, the search rule is given the following steps:
\begin{enumerate}
 \item Divide the board in theoretical squares of size $N_{\mbox{\tiny{y}}}\times N_{\mbox{\tiny{y}}}$ (see Fig. \ref{fig4}a)
 \item Click in the upper right corner of each subdivision. Start with those squares whose upper right cell has
 more neighbors and continue with those in the borders. This reduces the length of the extensive phase on average as corners that are farther from 
 the border are more likely to contain a yellow cell.
 \item Once a yellow cell is found, visit consecutive squares in a given direction (horizontal in Fig. \ref{fig4}a for $N_{\mbox{\tiny{y}}}=5$) until finding a black position. Then, 
 if the number of yellow cells in the row is lower than $N_{\mbox{\tiny{y}}}$, complete it.
 \item Repeat the same operation in the other direction starting from one already known yellow cell.
 \item Once the neighborhood of the target has been delimited, move inside it until finding the target.
\end{enumerate}

In the particular case of $N_{\mbox{\tiny{y}}}=5$, the average number of movements before target detection following this strategy is $19.03$ ($10^4$ realizations) and it is always lower
than $42$. In addition the extensive phase has a duration of $5.90$ clicks on average, which is about $50\%$ lower than the experimental result.  This improvement
is much higher than the one observed for the intensive phase, which can be optimized by players once they are provided with initial informaition
about the landscape (see gray and red bars in Fig.~\ref{fig-comp}b, c for a comparison). In real human scenarios, this result suggests that efforts put into optimizing the extensive phase may pay off more than equivalent efforts to 
optimize the intensive phase.

Applying this optimal strategy to many sizes of the yellow region (Fig. \ref{fig4}b) we observe that the tradeoff between finding the neighborhood of the target (yellow diamonds in Fig. \ref{fig4}b)
and finding the target inside it (blue circles in Fig. \ref{fig4}b) balances at intermediate values of $N_{\mbox{\tiny{y}}}$. Following theoretical results for blind experiments, 
analytical expressions can be obtained 
for the mean number of movements during both phases and therefore for the optimal interaction range. The mean number of 
clicks during the intensive phase is $N_{\mbox{\tiny{y}}}^{2}/2$ as the target can be in any cell with the same probability (green dashed 
line in Fig. \ref{fig4}b) (we only consider the lower bound obtained for blind experiments since this optimal protocol minimizes the number of
erroneous movements). To obtain the mean number
of movements in the extensive stage, we assume that the upper right corner of each subdivision of the board (Fig.~\ref{fig4}a) is equally likely to have a yellow cell.
Therefore, the number of steps is given by $N^{2}/2N_{\mbox{\tiny{y}}}^{2}$. This is not completely true, as cells close to the border
have a lower probability of being yellow, but it is a good approximation (black dashed 
line in Fig. \ref{fig4}b fitting yellow diamonds). At the optimal interaction range both functions intersect, which gives $N_{\mbox{\tiny{y}}}^{\mbox{\tiny{opt}}}=\sqrt{N}=4.47$
for our experimental setup with $N=20$. 
This result is in excellent agreement with the value obtained from the experiments (Fig. \ref{fig4}c) and suggests, together with the theoretical approximation,
that the optimal interaction range is independent of the searching strategy. This result suggests the possibility of using this theoretical framework
to predict the optimal size of the neighborhood of the target in more general scenarios.

\subsection*{Anticipating the optimal range of interaction for random neighborhoods.}
\begin{figure}[ht]
\centering
 \includegraphics[width=0.48\textwidth]{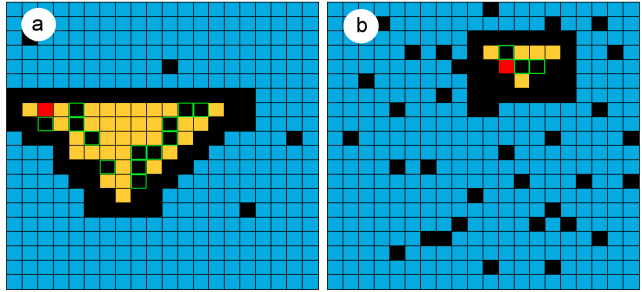}
\caption{\label{random-neigh} Construction of random information neighborhoods starting from triangles of different size. Black cells highlighted in green belonged to the original triangle
and have been removed in the randomization process. They are used here to indicate the original layout of the board but they are not shown to the player.}
\end{figure}

In this section we allow the target to adopt different sizes and random shapes across rounds. In order to facilitate the formulation of theoretical predictions,
the neighborhood is built starting from a triangle of varying base $b_y$ (see Methods for a detailed description and Fig.~\ref{random-neigh}) where
the target is embedded. Then, the region is randomized by turning $30\%$ of the cells black. In this way, we implement random neighborhoods that vary in form and size from round to round but
with an underlying fixed pattern. Before starting the game, players know that the neighborhood has now a varying form and size (Fig.~\ref{random-neigh}), but they are given no information about the way it is constructed.

\begin{figure}[ht]
\centering
 \includegraphics[width=0.48\textwidth]{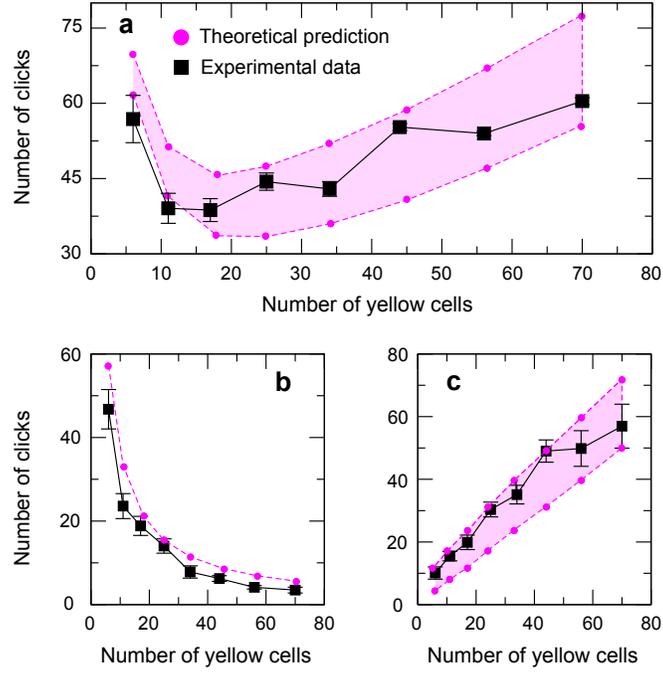}
\caption{\label{opt-line}  Prediction of the optimal size of random neighborhoods . a) Total number of movements, b) extensive phase, c) intensive stage. Black squares correspond to experimental data and magenta circles
to theoretical predictions. Dashed lines are interpolations and the error bars represent the standard error, when not shown they are smaller than the size of the square.}
\end{figure}

The optimal {\it interaction range} can be evaluated from an independent estimation of the number of movements needed in the extensive and the intensive phases. The length of the extensive phase
is obtained following the same steps used for square neighborhoods; the probability of finding a yellow cell in the $i-th$ movement is given by Eq.~(\ref{eq:extensive})
from where the mean length of the extensive phase is obtained using Eq.\ref{eq:limits-anal}.
This quantity is shown by the magenta circles in Fig.~\ref{opt-line}b. To approximate the number of movements used in the 
intensive phase, which will give us the optimal interaction range 
we used the underlying triangle shape of the neighborhood of the target. This calculation
provides lower and upper bounds for the average duration of the intensive phase. The lower bound is obtained assuming that all the cells from the original
target have the same probability of being visited
but all the cells that do not belong to it will never be clicked. The total number of cells that form this original triangle is $(b_y/2+0.5)^2$ and since all the cells
can be visited with the same probability, the lower limit for the length of the intensive phase is given by $\frac{(b_y/2+0.5)^2}{2}$. The upper limit is obtained assuming that the first cells that do not
belong to the triangle in each direction also have a non-zero probability of being visited. This results in an upper bound for the length of the intensive phase given by $\frac{[(b_y+4)/2+0.5]^2}{2}$. 
Both limits are shown by the magenta circles in Fig.~\ref{opt-line}c.
Finally, the total number of movements, i.e., the sum of the extensive and the intensive phase, is shown by the magenta circles in Fig.~\ref{opt-line}a, with an estimated optimal neighborhood size in between $18$ and $25$ yellow cells.
It is important to note the difference between the optimal interaction range for random and square neighborhoods, which shows the non triviality of predicting optimal interaction
ranges for different geometries.

We tested these a priori predictions with a series of experiments using an experimental setup with neighborhoods that consist of $5$, $16$, $33$, $55$ and $69$ cells (plus the target red cell, see Supplementary Table II for a distribution
of the number of rounds with each size). 
$301$ rounds were analyzed and the observed mean number of clicks is shown in Fig.~\ref{opt-line}a. We also split each round into the extensive and intensive phases and the results are shown in panels b and c of Fig.~\ref{opt-line}. 
The good agreement between the predicted values and the results obtained with the experiments shows the robustness of the theoretical approach developed in simpler scenarios.

\section*{Discussion}

We have developed a novel approach to study human 
search problems by building a simple game that can be accessed online. This approach facilitates the collection of large and clean experimental datasets.
By combining data analysis with probabilistic calculations and numerical simulations 
of existing and new models, it is possible to obtain a deeper understanding of how humans approach simple search tasks and how their strategies 
differ from optimal patterns. 

A comprehensive analysis of the trajectories on the board of the game (length of the displacements and turning angles) shows 
that players follow strategies consisting of two modes. The detection of cues about the location of the target
triggers  an area-restricted search mainly characterized by shorter movements on average \cite{hassell1978dynamics,curio2012ethology}.
In the context of existing studies, these processes are usually modeled by composite random walks that consist of an extensive phase and an intensive one. 
In the particular instance of animal foraging, the latter is triggered by encountering a food item and is characterized by shorter steps and larger turning angles (relative to the extensive mode)
\cite{MEE3:MEE312412,MendezChap6,benhamou1992efficiency,MoralesEtAl2004.Extracting.more}.  Our findings show that the duration of the search is minimal
when the cues extend over intermediate spatial scales as compared to the system size. The tradeoff between locating a cue and finding the target among the cues is balanced, which
results in faster searches. Although this result seems to be robust against changes in the total system size, considering larger landscapes could offer a richer phenomenology in the analysis of the trajectories on the board
as well as in the features of both phases.

In the simplest scenario studied here, in which no information is given about the size of the neighborhood of the target,
developing a systematic searching rule as opposed to following a stochastic trajectory does not provide a significant advantage.
A systematic scan of the environment usually provides higher efficiencies by minimizing the probability of revisiting a certain region.
In this setup, however, cells remain open once they are visited, providing players with a 
perfect memory about the history of their movements. As a consequence, neither random nor systematic players click more than once on a cell, regions are not revisited,
and both protocols offer equivalent results.
This scenario however changes when some information about the nature of the target is provided to the players. In that case an optimal systematic strategy can be constructed
based on this information. Interestingly, our data show that one of these optimal strategies was developed by a particular player who repeatedly played
several rounds in the same landscape. 
This result opens the door to explore a broad range of questions at the interface between landscape variability, the searcher's memory, and learning abilities, which has
recently become an important topic in movement ecology \cite{fagan2013spatial}. Most animals do not follow completely random strategies, 
but combine this stochastic component with spatial memory and learning \cite{merkle2014memory,boyer2014random,polansky2015elucidating}.
To investigate the importance of cognitive skills such as learning or memory in the development of optimal strategies, our approach could easily be extended to allow 
landscapes where the position of the target exhibits a certain degree of persistence across rounds of the game.  In addition, 
in order to compare how more complex decision-making processes come into play, it would be particularly interesting 
to compare the results presented here with the outcome of a new round of experiments in which players are not requested to find the target in the shortest possible time. 

In fact, we have shown that, when they have some knowledge about the landscape (size of the 
neighborhood of the target), players use the additional information obtained in each movement step to increase search efficiency.
In this scenario, the effect of the information gathered during the whole process has to be included in theoretical models to reproduce experimental results. 
Introducing a more realistic finite memory by allowing clicked cells to revert back to the unclicked state after some time arises as a future line of research.

More importantly, however, the excellent agreement between our experimental data and simple theoretical models suggest that 
this online-game based methodology could be applicable to address more complex scenarios. Energy budget related questions can be addressed by introducing a {\it metabolic} cost
that penalizes longer movements and {\it evolutionary} aspects of search problems may be 
addressed by allowing pairs of players to compete and selecting those using more efficient strategies. This would mimic environments where different individuals
compete for limited resources and could shed some light on the driving forces behind the evolution of optimal searching.
The effect of cooperative interactions among players on search efficiency could also be addressed. Many species forage in groups as opposed to individually. 
The methodology that has been presented here would facilitate, given a certain landscape, exploration of the level of confidence that players place
on movements performed by previous participants. Before every movement of the new player, the choice of previous searchers
at that same moment can be shown to the new player
to investigate whether and how much the current player trusts on previous participants. 
In addition, if the neighborhood of
the target is changed, or multiple targets are included, it would be possible
to explore the relationships between use of social information versus personal experience for tasks of increasing difficulty.
 Finally, in this study we have focused on the case of saltatory searches, in which the target can be detected only if the searcher lands on it. A next step
should consider the more general scenario of cruising searches, in which the target can be detected at any point of the displacements \cite{MendezChap6}. Such setup
would provide a higher flexibility in constructing more complex landscapes with different gradients of information that could allow the study of taxis-driven 
searches.

In summary, and in view of the large and exciting range of possibilities for future exploration, we expect that this general framework will complement purely theoretical efforts
to unveil the fundamental mechanisms that drive a wide variety of search scenarios.

\section*{Methods}

\textbf{Ethics statement.} The anonymity of all the participants was maintained during the whole experimental protocol. Participants accessed the game remotely through internet and non of their personal data was stored.
No ethical concerns are involved other than preserving the anonymity of participants. Informed consent was obtained from all subjects. The procedure was checked and approved by the
Committee of Ethics in Research of the University of the Balearic Islands, since the game was hosted in the web domain of one of its research institutes, the Institute for Cross-Disciplinary Physics and Complex Systems (IFISC)
The experiments were subsequently carried out in accordance with the
approved guidelines.

\subsection*{Fitting of the partial distributions of displacement lengths to gamma distributions in  informed  searches}

We showed that, for informed searches, the length of the displacements when players are given a priori information about the landscape follow a series of gamma distributions whose
probability density function is given by

\begin{equation}
 f(x;\alpha,\beta) = \frac{\beta^{-\alpha}{\rm e}^{-x/\beta}x^{-1+\alpha}}{\Gamma(\alpha)},
\end{equation}

where $\alpha$ and $\beta$ are real positive parameters. For known values of $\alpha$ and $\beta$, the mean value of the distribution can be obtained as $\alpha\beta$, the
variance as $\alpha\beta^{2}$ and the mode (the value that appears most often in the distribution) as $\beta(\alpha-1)$.  All the parameters shown in Table \ref{tabla-partidas}
were obtained using the maximum likelihood estimation. Results shown in the last row of Table \ref{tabla-partidas} correspond to a distribution that is a mixture
of all four component distributions.
Given the mean value and variance of these distributions, we can assume that they all have the same weight in the composition since all the subsets of the 
trajectory have the same length. The mixed distribution can be obtained as:

\begin{eqnarray}
 \mu_{\mbox{\tiny{mix}}} &=& \frac{1}{4}\sum_{i=1}^{4}\mu_i \\
 \sigma^{2}_{\mbox{\tiny{mix}}} &=&  \frac{1}{4}\sum_{i=1}^{4} \left(\mu_{i}^{2}+\sigma_{i}\right)-\mu_{\mbox{\tiny{mix}}}^{2}
\end{eqnarray}

\subsection*{Implementation of the random walk model for blind searches}\label{model:markov}
We have developed a minimalistic model based on composite random walks to understand the basic features of the search strategies used by
the players. We initialize the model from a random configuration of the board in which the target is placed in a random position inside a smaller square of lateral length $N_{\mbox{\tiny{y}}}$.
To mimic the experimental setup, we fix the size of the board so it has $20$ cells on each side and explore $N_{\mbox{\tiny{y}}}$ varying between $1$ and $20$. 
The searcher is placed in a random position of the board and the dynamics starts. The algorithm consists of the following steps:
\begin{enumerate}
 \item Obtain the probability of jumping from the current position, $i$ to the rest of the cells in the board $j$. This is given by
 the experimental jump length distributions, so $P_{ij}=\exp(-\lambda_\gamma r_{ij})/\lambda_{\gamma}$,
 where $\gamma\equiv\lbrace \mbox{in, ext} \rbrace$ and $r_{ij}$ is the distance between two cells.
 The two values of $\lambda$ are obtained from the experimental 
 data and define the extensive and the intensive phase: $1/\lambda_{\mbox{\tiny{int}}}=2.05$ and $1/\lambda_{\mbox{\tiny{ext}}}=3.70$.
 \item As in the game the player has perfect memory of previous moves, so the probability of jumping to already visited cells is set to zero. 
 \item If any of the visited cells belongs to the neighborhood of the target (yellow cell),
 then we multiply the probability of jumping to each of its unvisited neighbors by a bias factor $\eta=10^{3}$ whose effect  
 is to keep the searcher around the cues and avoid unrealistic escapes from them. The existence of such a bias is suggested by the distribution of turn angles shown 
 in Fig.~\ref{fig:blindexp}d that shows a high probability of returning to the yellow region when it is left.
 Our results are, however, independent of the numerical value 
 of this bias provided that it is strong enough to trap the searcher close to the yellow cells.
 \item Renormalize all the jumping probabilities so $\sum\limits_{j=1}^{N^2}P_{ij}=1$.
 \item Sort a uniform random number $u$ between $0$ and $1$ and move to a cell $k$ when $\sum\limits_{j=1}^{k}P_{ij}\geq u$.
\end{enumerate}
These steps are repeated until the target is found, then the number of movements is saved and the system restarted for a new realization.

\subsection*{Implementation of the random walk model for searches with initial information}\label{model:nonmarkov}
To introduce the effect of having initial information about the configuration of the landscape (size of the yellow region) we modify the random-walk model presented for blind searches.
Simulations are set as in the first model, starting from a $20\times 20$ cells board where the target is randomly placed inside a square region of lateral length $N_{\mbox{\tiny{y}}}=5$.
The position of this region is also random in the board and changes across realizations. 
The searcher is placed at an initial random position and the dynamics starts. The algorithm has two well differentiated parts for the intensive and the extensive phase:
\begin{itemize}
 \item Extensive phase:
\begin{enumerate}
 \item Obtain the distance from the current position of the searcher, $i$, to every other cell in the board, $j$,
 and assign a jumping probability, $P_{ij}$, by taking a random sample from the histogram in Fig.~\ref{fig3}a.
 \item As in the game the player has perfect memory of previous moves, so the probability of jumping to already visited cells is set to zero.
 \item For every cell $j$ in the board obtain all the possible $5\times 5$ squares to which it can belong. 
 If all of them have any open black cell, then set the probability of jumping to $j$ to zero. This step is skipped in the
 intermediate model where only the intensive phase is improved.
 \item Renormalize all the jumping probabilities so they sum one.
 \item Sort a uniform random number $u$ between $0$ and $1$ and move to a cell $k$ when $\sum\limits_{j=1}^{k}P_{ij}\geq u$.
\end{enumerate}
 \item Intensive phase, after the first yellow cell is hit:
 \begin{enumerate}
 \item Obtain all the possible neighborhoods of the target to which the first detected yellow cell can belong. 
 \item Count the number of open cells of both classes (black and yellow) in each of those possible neighborhoods of the target.
 \item Pick those $5\times5$ squares that include all the open yellow cells and none of the black ones.
 \item Set the probability of jumping to all other of the rest of the cells of the board to zero.
 \item From the histogram in the inset of Fig.~\ref{fig3}a, obtain the probability $P_{ij}$ of jumping to the cells that belong to the chosen $5\times5$ squares.
 \item Renormalize all the jumping probabilities so they sum one.
 \item Sort a uniform random number $u$ between $0$ and $1$ and move to a cell $k$ when $\sum\limits_{j=1}^{k}P_{ij}\geq u$.
 \end{enumerate}

\end{itemize}

% \bibliography{references}

\section*{Acknowledgements}

We acknowledge Ant\`{o}nia Tugores, Rub\'en Tolosa and Iharob al Asimi Espina for advice in the development of the experimental setup. We are also grateful to George W. Constable for useful discussions 
 and to Frederic Bartumeus for useful discussions and a critical reading of the manuscript. This work is funded by the Gordon and Betty Moore Foundation through Grant
GBMF2550.06 to RMG, Universitat de les Illes Balears through a 2015 Young Visiting Scholar grant to RMG, the  US National Science Foundation through grant ABI 1458748 to JMC and
Ministerio de Econom\'ia y Competitividad and Fondo Europeo de Desarrollo Regional through project CTM2015-66407-P (MINECO/FEDER) to CL.

\section*{Author contributions statement}
R.M-G conceived the study, implemented the experimental setup, and did the numerical simulations.
All the authors designed the experiments, analyzed and discussed the results and contributed to the writing and revision of the manuscript.

\section*{Additional information}
The authors declare no competing financial interests.

\begin{table}
\centering 
\begin{tabular}{| c || c | c | c | c | c |}
\hline
 \ \ Part \ \ & \ Mean \ & \ Variance \ & \ Mode \ & \ \ \ \ $\alpha$ \ \ \ \ & \ \ \ \  $\beta$ \ \ \ \ \\[0.5ex] \hline\hline 
1-5 & 8.48 & 22.56 & 5.83 & 3.19 & 2.66 \\
6-10 & 6.66 & 12.86 & 4.72 & 3.45 & 1.93 \\
11-15 & 6.09 & 9.36 & 4.56 & 3.97 & 1.54 \\
16 -- & 5.20 & 7.88 & 3.69 & 3.44 & 1.51 \\ \hline
Total & 7.07 & 16.10 & 4.80 & 3.11 & 2.28 \\
Mix & 6.61 & 16.02 & 4.19 & 2.73 & 2.42 \\
\hline
\end{tabular}
\caption{ Parameters obtained fitting the jump length distributions to gamma distributions. The extensive phase of the informed searches
is divided in four pieces and the partial distributions fitted to gamma distributions. Changes in the mean value show the non-stationarity of the process .} 
\label{tabla-partidas}
\end{table}

\end{document}